\documentclass{aa}  

\usepackage{graphicx}
\usepackage{txfonts}
\begin{document}

\title{Weighing the two stellar components of the Galactic Bulge}

   \author{M. Zoccali,
          \inst{1,2}
          E. Valenti,
          \inst{3}
          O. A. Gonzalez,
          \inst{4}
          }

\institute{Instituto de Astrof\'{i}sica, Pontificia Universidad Cat\'{o}lica de Chile, 
Av. Vicu\~{n}a Mackenna 4860, Santiago , Chile \\
    \email{mzoccali@astro.puc.cl}
\and
Millennium Institute of Astrophysics, Av. Vicu\~{n}a Mackenna 4860, 782-0436 Macul, 
Santiago, Chile 
 \and
European Southern Observatory, Karl Schwarzschild\--Stra\ss e 2, D\--85748 Garching bei 
M\"{u}nchen, Germany 
\and
UK Astronomy Technology Centre, Royal Observatory, Edinburgh, EH9 3HJ, UK
}
   \date{}

 
  \abstract
      {Recent spectroscopic surveys of the Galactic bulge have unambiguously shown that the
       bulge contains two main components, that are best separated in their iron content, but also
       differ in spatial distribution, kinematics, and abundance ratios. The so-called metal poor
       component peaks at [Fe/H]$\sim$$-0.4$, while the metal rich one peaks at [Fe/H]$\sim$+0.3.
       The total metallicity distribution function is therefore bimodal, with a dip at [Fe/H]$\sim$0.
       The relative fraction of the two components changes significantly across the bulge area.
      }
   {We provide, for the first time, the fractional contribution of the metal poor and metal rich stars
   to the stellar mass budget of the Galactic bulge, and its variation across the bulge area.}
   {This  result  follows from  the  combination of the  stellar mass  profile  derived
     empirically by \citet{valenti+16}  from VISTA Variables in the  V\'\i a L\'actea
     data, with the relative fraction of metal  poor and metal rich stars, across the
     bulge  area,  derived  by  \citet{zoccali+17}   from  the  GIRAFFE  Inner  Bulge
     spectroscopic Survey.  }
   {We find  that metal poor stars  make up 48$\%$ of  the total stellar mass  of the
     bulge, within the region $|l|$$<$10, $|b|$$<$9.5, with the remaining 52$\%$ made
     up of  metal rich  stars. The  latter dominate the  mass budget  at intermediate
     latitudes $|b|$$\sim$4, but become marginal  in the outer bulge ($|b|$$>$8). The
     metal poor component is more axisymmetric than  the metal rich one, and it is at
     least comparable, and possibly slightly dominant in the inner few degrees.  
       As a result, the metal poor component, which does not follow the main bar, is
       not marginal in terms of the total mass budget as previously thought, and this
       new observational evidence must be included in bulge models.
     While
     the total radial velocity dispersion has  a trend that follows the total stellar
     mass, when we examine the velocity dispersion of each component individually, we
     find that metal poor stars have higher  velocity dispersion where they make up a
     smaller fraction of the stellar mass, and  viceversa. This is due to
     the kinematical and spatial distribution  of the two metallicity component being
     significantly different, as already discussed in the literature.  }
   {}

   \authorrunning{Zoccali et al.}  \maketitle

\section{Introduction}

The Galactic bulge is qualitatively known to be the massive, old component of 
the Milky Way, therefore a very important region to study in order to understand the
early process(es) leading to the formation of our Galaxy. How massive, and how old,
however, is still a matter of debate.

The age of bulge  stars has been the most controversial topic in  the last few years.
In  fact, studies  of the  color-magnitude diagram  (CMD) find  that bulge  stars are
mostly as  old as $\sim$10 Gyr  \citep[e.g.,][]{ortolani+95, zoccali+03, clarkson+08,
  valenti+13}.   On  the  contrary,  spectroscopic measurements  of  individual  main
sequence turnoff stars, during a microlensing event that made them observable at high
spectral resolution,  find that only metal  poor (MP) stars are  uniformly old, while
metal rich (MR) ones span a wide range of ages, most of them being younger than 6 Gyr
\citep{bensby+17}. Compatible  results have been found  by \citet{schultheis+17} from
spectroscopic masses (therefore ages) for red giant branch (RGB) stars based on C and
N abundances calibrated  agains asteroseismic data.  Some effects  have been proposed
to reconcile at least part of the discrepancy, such as the difficulty in discerning a 
MR, young population from a MP, old one in the CMD because of the age-metallicity 
degeneracy \citep{haywood+16}; or  a possible  overabundance  of helium  for the  MR
population \citep{nataf+16}. In particular, \citep{haywood+16} presented the analysis
of a deep Hubble Space Telescope (HST) CMD  of a bulge field and concluded that there
must be  a wide range  of ages  in the bulge  to simultaneously reproduce  the narrow
width of the observed main-sequence turn-off and the spectroscopic metallicity spread
under any  reasonable age-metallicity  relation. In  agreement with  this conclusion,
Bernard et al.  (2018)  calculated the star formation history of  the bulge using the
same HST field and concluded that over $\rm80\%$  of the stars formed more than 8 Gyr
ago, but that a significant fraction of the super-solar metallicity stars are younger
than this age.

While the age distribution of the bulge, as well as its spatial variation, remains to
be fully understood, different spectroscopic studies of large samples of bulge giants
have agreed on the fact that bulge MP and MR stars have different properties.

The first evidence for a bimodality in the metallicity distribution function (MDF) of
a  sample   of  $\sim$400  K  giants   in  Baade's  Window  has   been  presented  by
\citet{hill+11}.  The bimodality was  not so striking in their MDF,  but the case for
the existence of  two metallicity population was reinforced by  a different behaviour
of  MP/MR stars  in the  [Mg/Fe] versus  [Fe/H] plane,  and by  different kinematical
properties, discussed  in the  companion paper  by \citet{babusiaux+10}.  The latter
detected a different trend of the velocity dispersion versus metallicity, in Baade's
Window   and    in   two   other    fields   at   $b$=$-6$   and    $b$=$-12$,   from
\citet{zoccali+08}.  \citet{babusiaux+10}  also  combined the  radial  velocity  from
spectra  with proper  motions from  the OGLE  II survey  \citep{sumi04}, in  order to
derive  3D velocities.  This allowed  them  to detect  that  MP stars  have a  rather
isotropic orbit distribution, typical of axisymmetric spheroids, while MR stars show
elongated orbits, characteristics of Galactic bars. This was illustrated by the trend
of the  vertex deviation versus [Fe/H],  shown in their Fig.~3.   Updated versions of
the same  plot are presented  in \citet{babusiaux16}, together  with a review  of the
kinematics of bulge stars.

Large spectroscopic  surveys of  bulge giants  in several fields,  such as  the ARGOS
survey     \citep{freeman+13},     the     Gaia-ESO     survey     \citep{gilmore+12,
rojas-arriagada+14}, and  the GIBS ESO Large  Programme \citep{zoccali+14} confirmed
that the MDF is clearly bimodal in every direction probed so far, and that MP and MR
stars have  different spatial distribution  and different radial velocity dispersion
($\sigma$).  

Specifically,  MP  stars  peak  at   [Fe/H]$\sim$$-0.4$,  and  do  not  extend  below
[Fe/H]$<$$-1$. They have  a more axisymmetric spatial distribution than  the MR ones,
and their  radial velocity dispersion  shows a  mild spatial variation,  ranging from
$\sim$80 km/s at  $b$=$-8$ to $\sim$120 km/s  at $b$=$-1$. Instead, MR  stars peak at
[Fe/H]=+0.3,  reaching (barely)  metallicities as  high as  [Fe/H]=+0.7. They  show a
remarkably  boxy distribution  in the  plane of  the sky,  and their  radial velocity
dispersion has a larger spatial gradient, from $\sim$50 km/s at $b$=$-8$, to 140 km/s
at $b$=$-1$.  The numbers quoted here are from the GIBS programme \citep{zoccali+17},
although fully consistent results are   obtained      from       Gaia-ESO       data
\citep{rojas-arriagada+17}. Independent evidence for a slower rotation of the MP stars, 
compared to the MR ones, also comes from Clarkson et al. (2018, ApJ {\it in press}). Results 
from the  ARGOS survey \citep{ness+13a, ness+13b}
are somewhat different because they are based on a bigger target selection box in the
CMD,  therefore  including  a  larger  contribution  from  the  thick  disk  and  the
halo. \citet{ness+13a}  fit the  MDF with 5  components, with the  two most  MR being
qualitatively  compatible with  the  MP and  MR components  from  GIBS and  Gaia-ESO,
respectively. 

An additional confirmation of the fact that  MP and MR stars have a different spatial
distribution  in   the  Galactic  bulge   came  from  \citet{ness+12}.   The  overall
distribution of red  clump (RC) stars in  the bulge is known to  follow a boxy/peanut
(B/P) shaped structure \citep{wegg+13}. Such structures  are known to be the products
of  bars  suffering  vertical  instabilities  \citep{patsis+02,athanassoula05}.  This
characteristic  shape was  first  identified by  a  double RC  seen  along the  bulge
projected    minor-axis   for    latitudes    $b$$>$$-5$   \citep{mwz10,    nataf+10,
  saito+11}. \citet{ness+12} used metallicities measured in three fields of the ARGOS
survey, to show that the double red-clump  distribution is traced by the MR stars and
not by  the MP ones.  The result  was later confirmed  by \citet{rojas-arriagada+14},
using the same approach, applied to data from the Gaia-ESO survey.

Similarly,  \citet{gran+16}  identified RR  Lyrae  variables  in the  VVV  catalogues
\citep{minniti+10,  saito+12} for  the  outer bulge  $-8<b<-10.3$  and confirmed,  by
measuring their  distances, that  RR Lyraes  do not follow  the strong  B/P structure
traced by the MR  RC stars.  The spatial distribution of RR  Lyrae variables across a
wide   bulge    area   was   independently   analysed    by   \citet{pietrukowicz+12,
  pietrukowicz+15},   using  OGLE   III  photometry   \citep{soszynski+11},  and   by
\citet{dekany+13}, combining OGLE  III and VVV photometry.  Both groups  find that RR
Lyrae  variables  do not  follow  the  main  Galactic  bar\footnote{A review  of  the
  observational proofs of the presence of a strong bar in the Milky Way is beyond the
  purpose of the present paper.  We refer  the interested reader to the recent review
  paper on the  bulge 3D structure by \citet{zoccali+16}}, although  they disagree on
their spatial distribution being slightly elongated in the same direction as the bar,
or completely axisymmetric, respectively. Support to the result that RR Lyrae trace a
different component comes from the fact that they seem to show no rotation 
\cite{kunder+16}.

 There are,  in  the literature,  two  sources of  confusion  concerning the  MP
  component  in the  Galactic bulge.  One is  that, if  it has  a spheroidal  spatial
  distribution, then it must have a {\it  classical bulge} origin, i.e., it must have
  formed  from  violent merging  of  substructures,  either  gas clumps  or  external
  building blocks.  Some  recent models argue that this is  not necessarily the case.
  \citep{debattista+17} showed, using a simulation with star-formation, that the same
  behaviour can be explained  by the redistribution of populations by  the bar on the
  basis of  their disc  kinematics at the  time of bar  formation. As  a consequence,
  younger, MR  populations would  display a  stronger bar, and  a more  prominent B/P
  shape, than the older, MP ones.  A similar conclusion was presented by Fragkoudi et
  al.  (2017) using numerical simulations composed of distinct thin and thick discs.

The second misconception is that the MP, spheroidal component represents a minor
  contribution to the total stellar mass of the bulge. This idea is based on the fact
  that, in Baade's Window, by far the most studied region of the bulge, the MDF shows
  a strong  peak at  supersolar metallicity (the  MR component) with  a sort  of tail
  extending to [Fe/H]=$-1.0$ (the MP component). The Baade's Window MDF has been used
  as representative of the whole bulge by  several authors, when comparing it to both
  models and  extragalactic bulges. Only very  recently it was demonstrated  that the
  bulge MDF shows large variations across the  bulge area, and that MP stars are very
  abundant in the inner few degrees of the  Milky Way. We here quantify for the first
  time the contribution of  the MP component to the total stellar  mass of the bulge,
  demonstrating that  it makes up  about half of  it.  Formation models  must include
  this component, in the correct amount  and with the correct spatial and kinematical
  properties, when trying to explain the origin of the Milky Way bulge.

\section{The Bulge Projected Mass Distribution}

By  using star  counts from  the  VVV survey,  in \citet{valenti+16}  we derived  the
projected  stellar density  profile  of the  bulge  within the  sky  area defined  by
$|l|$$<$10, $|b|$$<$9.5. We counted RC stars in PSF-fitting photometric catalogues of
$J$  and $Ks$  VVV images,  corrected  for completeness  and interstellar  extinction
\citep{gonzalez+12}. This was done under the important assumption that RC stars trace
the global stellar population.

In the same paper, we derived an  empirical conversion between the number of RC stars
and the  total stellar mass.  This conversion was obtained  using the bulge  IMF from
near-IR NICMOS  data by \cite{zoccali+00}, later  combined with near-IR SOFI  data to
derive a  complete LF, for the  same field \citep{zoccali+03}. The  complete LF gives
the number of RC stars, while the total  stellar mass is the integral of the IMF both
measured in  the same field.   With this  method we converted  the number-of-RC-stars
profile  into a  stellar mass  profile, and,  by integrating  it, finally  derived an
empirical  estimate of  the bulge  total stellar  mass, resulting  in $2\times10^{10}
M_\odot$ with an uncertainty of 15\%.

\begin{figure}
\centering
\includegraphics[width=9cm,angle=0]{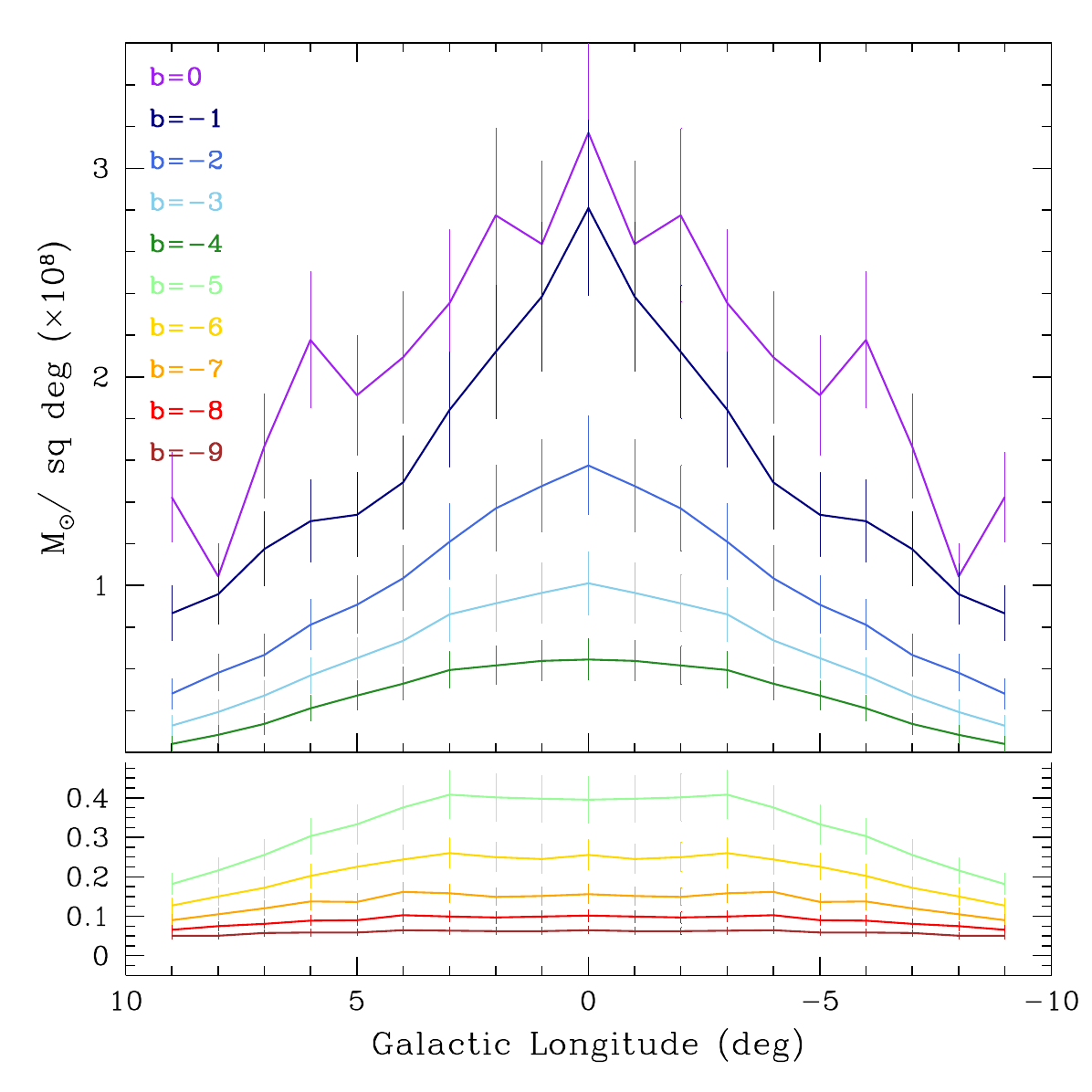}
   \caption{Stellar mass profile of the Galactic bulge, as a function of longitude,
for different latitudes. The curves at low latitudes are shown on a different vertical
scale, to increase visibility. This is a new version of Fig.~5 in \citet{valenti+16}
where we have imposed symmetry about the Galactic plane and about the bulge projected 
minor axis ($l=0$). Statistical uncertainties, coming from the Poisson fluctuation
  on the number of RC stars are $\sim$1\%. Error bars of 15\% of the data points are
  shown here, in order to account for the systematics, such as the IMF slope, the disk
  contamination and the adopted size of the RC box; see \citet{valenti+16} for a discussion.
}
\label{fig:mass}
\end{figure}

There  are  two important  assumptions  in  this approach.   The  first  is that  the
measurements  of  the  total  LF  and  the  IMF for  the  bulge  are  correct  as  in
\citet{zoccali+03}  and \citet{zoccali+00},  respectively.  Because  all the  numbers
needed for the calculation are given, new  assumptions for the bulge LF and IMF allow
one to derivate a new value for the  bulge mass.  For example, a new IMF was measured
by \citet{calamida+15}, based on optical HST data, with proper motion decontamination
from  foreground  disk  stars.   Their  best   fitting  power  law  has  a  slope  of
$\alpha$=$-2.41$ for masses in the range 1$-$0.56 $M_{\odot}$ and $\alpha$=$-1.25$ in
the   range  0.56$-$0.15   $M_{\odot}$.    By   using  this   IMF   instead  of   the
\citet{zoccali+00} one, and  mantaining the assumptions for the brown  dwarfs and the
remnants, the total bulge stellar mass is $2.1\times 10^{10}$ $M_{\odot}$, consistent
with the value derived in \citet{valenti+16}.

\begin{figure*}[ht]
\centering
\includegraphics[angle=0,width=8cm]{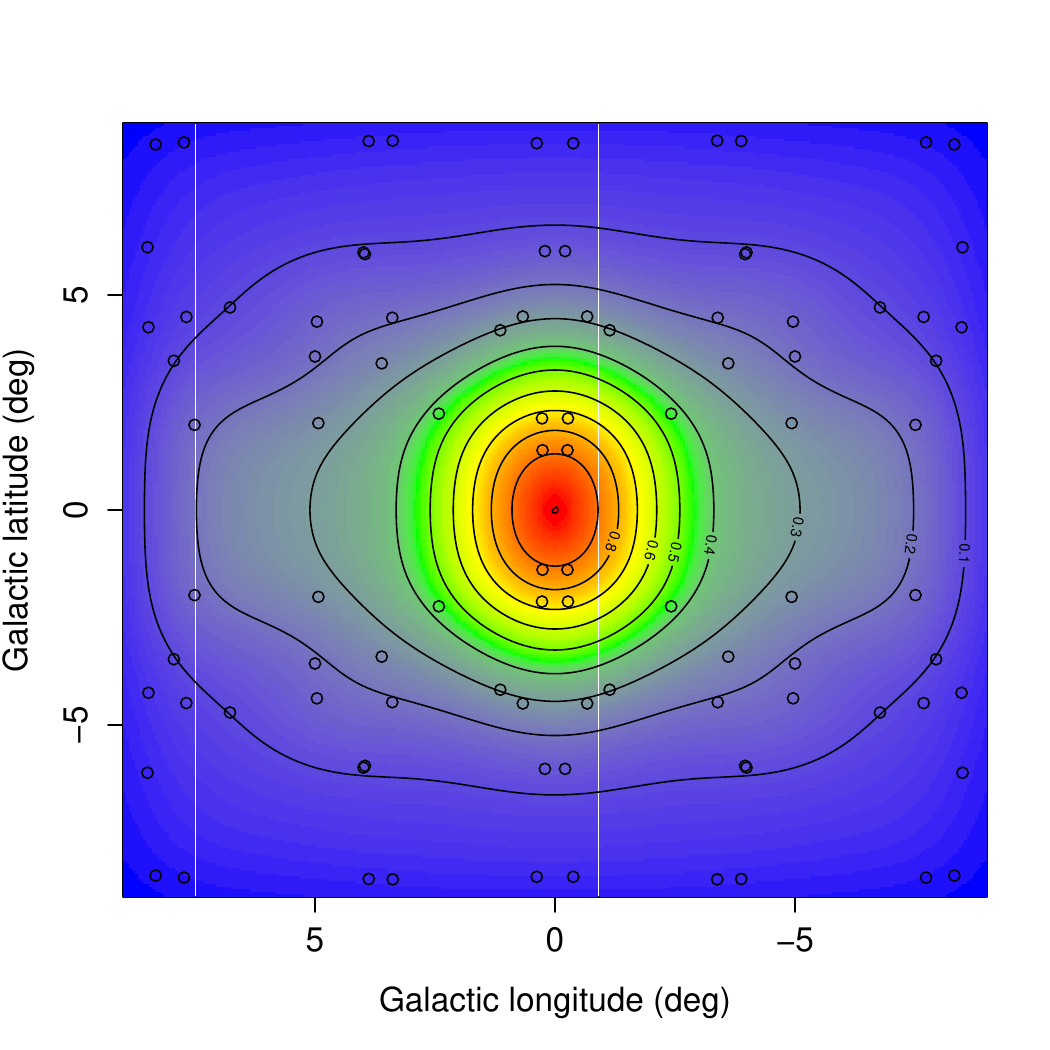}
\includegraphics[angle=0,width=8cm]{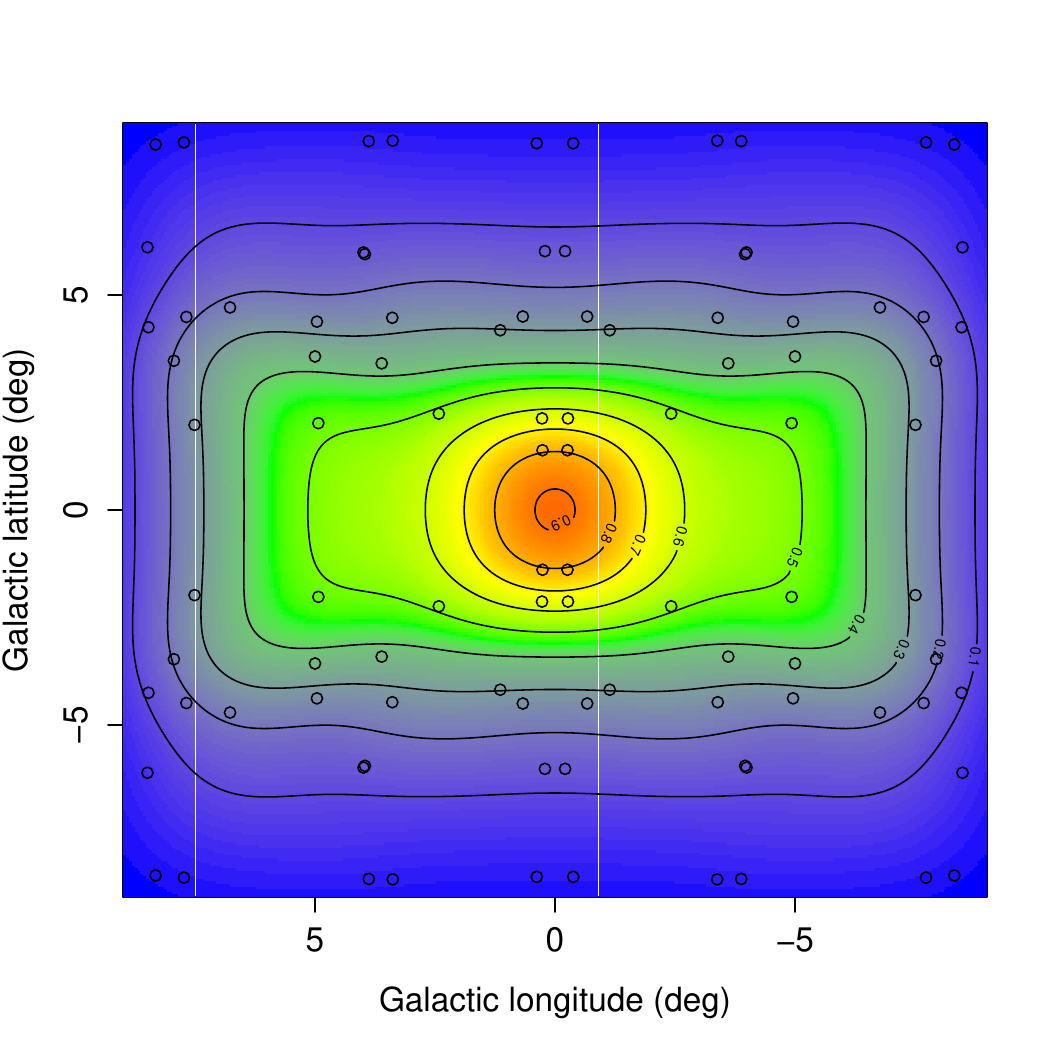}
\caption{Number density maps of the MP (left) and MR (right) components, obtained
  by convolving the MP/MR relative fraction of stars from the metallicity distribution
function, and the global stellar density map from \citet{valenti+16}. Open circles 
  show the grid of fields used in the interpolation, resulting from folding the 
  measurements in the GIBS fields with respect to both Galactic latitude and
  longitude. The maps are normalized to the same color scale. The figure is 
  reproduced from \citet{zoccali+17}; see the original paper for details.}
\label{fig:mappe-gibs}
\end{figure*}

The second assumption  is the fact that when  we count stars in the  RC+RGB region of
the bulge CMD, we are indeed counting only bulge stars. The normalization between the
stars  counted  in the  whole  VVV  area  and the  bulge  LF  in  the SOFI  field  in
\citet{zoccali+03} was  done in a  box of  $\pm 1.5$ mag  across the RC  peak.  These
stars are used  as tracers of the  underlying bulge stellar density, which  is a safe
hypothesis  unless  there is  another  population,  along  the  line of  sight,  that
contributes to the counts in this range\footnote{Notice that our counts are corrected
  for  an estimated  18\%  contamination by  foreground disk  RC  stars. A  different
  percent  contribution  can be  assumed  by  other  authors, and  the  corresponding
  correction applied.}.   In addition, the  distance limits within which  the stellar
mass  is measured  are formally  fixed  by this  $\pm  1.5$ mag  interval across  the
RC. However, because this magnitude range  corresponds to distance limits much larger
than the bulge extension, they might include other components. In order to assess the
impact of the  adopted normalization box on  the final estimation of  the bulge total
stellar mass,  we repeated the calculation  counting stars within $\pm  1.0$ mag, and
within $\pm 0.5$  mag, across the RC  peak.  The resulting bulge  mass is $1.93\times
10^{10}$ and $2.22\times 10^{10}$ $M_\odot$,  respectively, consistent to the $2.0\pm
0.3  \times 10^{10}$  $M_\odot$  given  in \citet{valenti+16},  and  well within  the
estimated uncertainty.  In what  follows we will use the latter  value as total bulge
stellar mass,  when needed.  However here  we determine the relative  contribution of
the MP  and MR component  to the  total stellar mass,  which is independent  from the
value adopted for the latter.

In order to derive  the density and mass profile for the bulge  MP and MR components,
separately, we combine  Valenti's results with the fractional density  maps of MP and
MR  stars obtained  by  \citet{zoccali+17} from  the GIBS  program.  These maps  were
derived by imposing 4-fold spatial symmetry,  about the Galactic plane, and about the
projected minor  axis. Given that {\it  i)} the stellar density  is mostly symmetric;
and {\it  ii)} we use  symmetric maps  from GIBS, we  produce here a  4-fold averaged
version  of  the   mass  profile  from  \citet[their   Fig.5]{valenti+16},  shown  in
Fig.~\ref{fig:mass}.  The  figure shows a smooth  increase of the bulge  stellar mass
when  moving towards  the Galactic  center,  with the  presence of  a boxy  structure
evident at  latitudes $b$=$-4$,$-5$,$-6$. The zig-zag  profile at $b$=0 is  due to the 
observational  complications of this region  with much larger  extinction (therefore  
more  uncertain  differential extinction  correction), lower completeness and stronger 
foreground disk contamination.

\section{The Projected density distribution of MP and MR stars}

Figure~\ref{fig:mass} shows the projected profile of the total stellar mass contained
in the bulge. As reviewed in  the introduction, however, all the recent spectroscopic
measurements, unambiguously show that the bulge contains two stellar components. They
are best separated in the metallicity ([Fe/H]) distribution, but differ also in their
spatial   distribution,   kinematical   properties,  and   alpha-element   abundances.

\citet{zoccali+17} combined the  relative fraction of the  two populations, extracted
from the GIBS metallicity distribution function in  each of 26 bulge fields, with the
stellar  density map  in \citet{valenti+16}.  In this  way they  derived the  stellar
number density of each of the MP and MR component,  in the 26 GIBS fields.  By means of a 2D
interpolation,  they obtained  a  stellar density  map  for  the MP  and  for the  MR
component,  reproduced  here  in  Fig.~\ref{fig:mappe-gibs}.   The  maps  are  4-fold
symmetric by  construction (about the Galactic  plane and about the  $l=0$ axis), and
they clearly show that the MP component is axysimmetric,  i.e., circular when projected
in the plane of the sky, while the MR one is markedly
boxy. Because  the maps are  normalized to the same  maximum density, they  also show
that the MP  component reaches a slightly  higher density in the  central region.  We
emphasize  that the  innermost GIBS  field  is at  ($l,b$)=($-0.26,-1.4$), where  the
fraction of MP stars with respect to  the total is 53\%. Therefore, the overabundance
of MP stars in  the center is marginal. It is however very  interesting that MP stars
are {\it at least as abundant as} MR ones in the Milky Way central region: this was a
new and unexpected result of the GIBS project.

\section{The mass fraction of MP and MR stars}

The combination of  the GIBS maps presented  in the previous section  and the stellar
mass profile in Fig.~\ref{fig:mass} allows us to derive mass profiles for each of the
MP and MR  component, individually, that we list in  the online Table~\ref{tab:mass}.
Conceptually, these carry the same information  on the spatial distribution of MP and
MR stars as the maps in  Fig.~\ref{fig:mappe-gibs}.  Converting these density maps in
units of stellar  mass, however, allows us  to integrate them across  the whole bulge
area, hence deriving the percent contribution of the MP and MR component to the total
stellar mass of the bulge.

\begin{figure}
\centering
\includegraphics[width=9cm,angle=0]{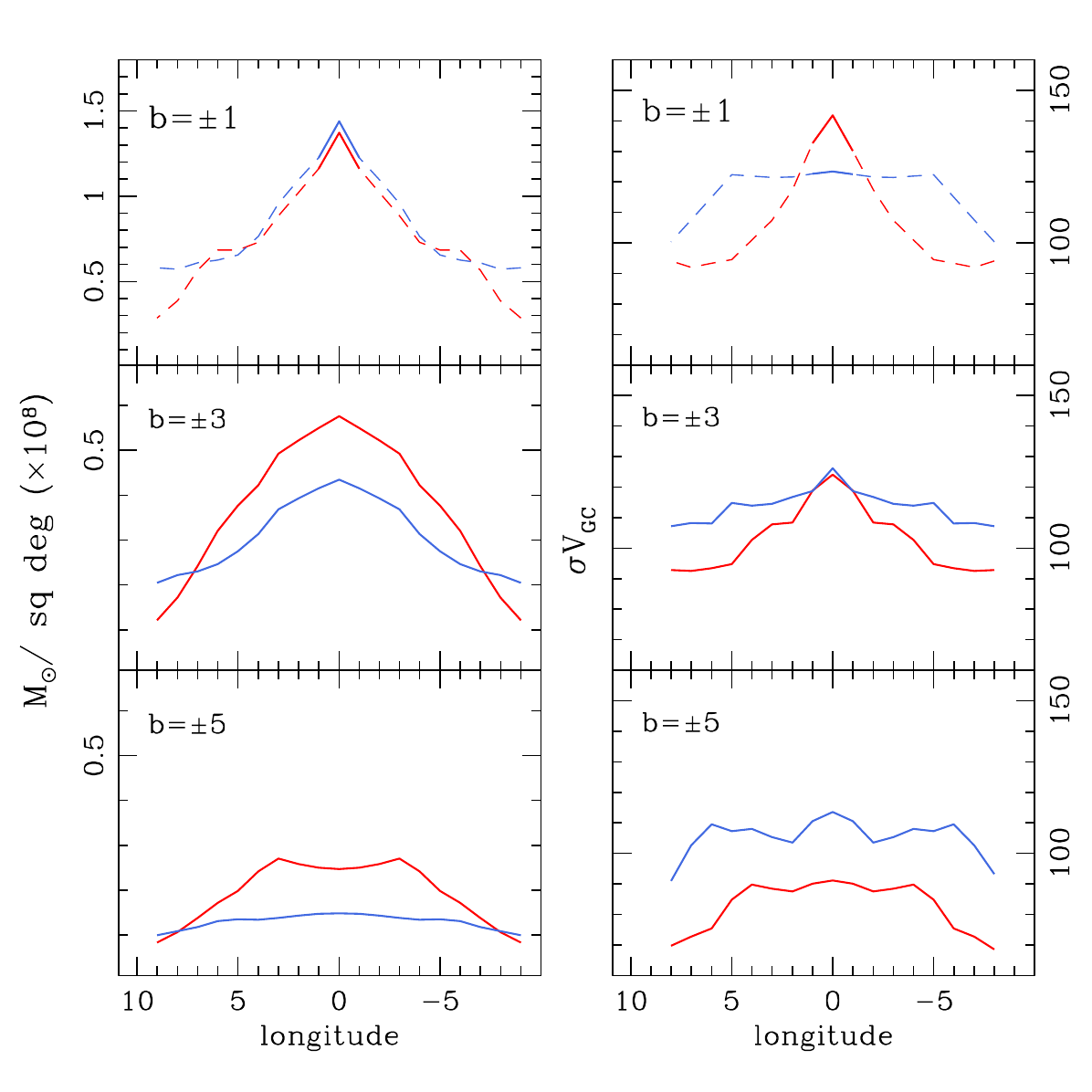}
   \caption{{\it Left panels:} stellar mass profile as a function of longitude, for
three different latitudes, for MP (blue) and MR (red) stars, respectively. {\it Right
panels:} same thing for the galactocentric radial velocity profile. At $b$=$\pm1$ the
curves are dashed, except at $l$=$0$, to remind the reader that they are constrained
by a single field at ($l,b$)$\sim$($0,-1$), and -only through the interpolation- by the 
fields at $b$=$-2$). }
\label{fig:mass-sigma}
\end{figure}

The left  panels of Fig.~\ref{fig:mass-sigma}  show the mass profile  with longitude,
plotted at three representative latitudes, for the MP (blue) and MR (red) components,
individually.  These are  the curves shown in navy blue  ($b$=$-1$), azure ($b$=$-3$)
and light green ($b$=$-5$) in Fig.~\ref{fig:mass} now splitted according to the MP/MR
fraction in each point, as given by the GIBS maps.

As expected,  MR stars  are more  abundant than MP  stars in  the bulge  at latitudes
$|b|$$>$$3$ (bottom and  central panels).  MR stars also show  the boxy structure, at
$b$=$\pm5$. They  are slighly underabundant  at $b$=$-1$,  where, we recall,  we have
data only  for a single  field at $l$$\sim$$0$, therefore  we plot the  profiles with
dashed lines at other  longitudes. MP stars, on the other hand,  have a rather smooth
trend,  from  a  flat and  marginal  contribution  at  $|b|>5$  to a  strong  central
concentration as we  move closer to the  Galactic plane. The mass profile  of the two
components at $b$=$0$ is not shown, because it would be just an interpolation, due to
the lack of spectroscopic data at this latitude.

\begin{figure*}[ht]
\centering
\includegraphics[angle=0,width=8cm]{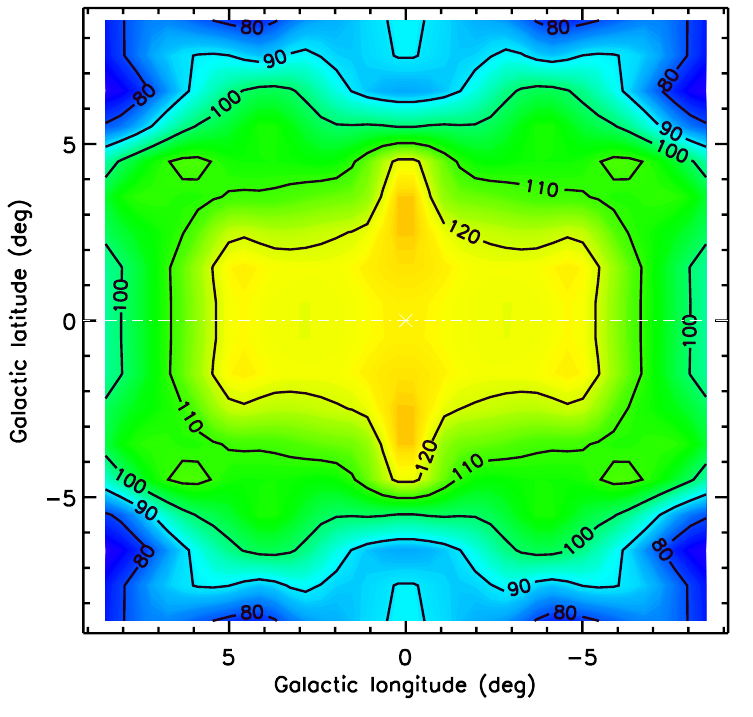}
\includegraphics[angle=0,width=8cm]{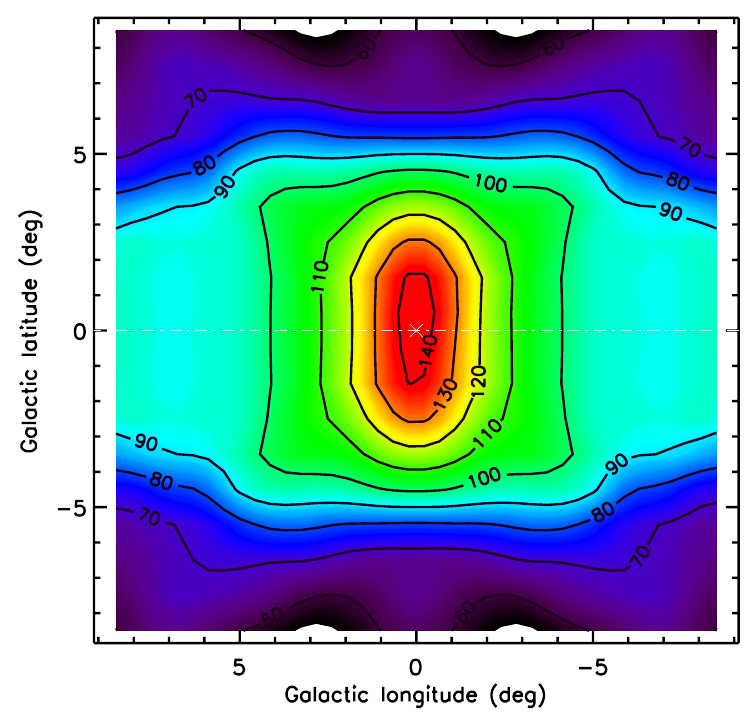}
\caption{Spatial trends of  the galactocentric radial velocity dispersion for the  MP (left) and
  MR  (right)  component,  respectively.  The  maps  are  the  results  of  a  linear
  interpolation between the measured value of the $\sigma$ for the two components, in
  the  26  GIBS  fields.   The  maps  are symmetric  in  longitude  and  latitude  by
  construction, and they have been normalized  to the same color scale. Contours show
  the value of $\sigma$, in km/s, in step of 10.  }
\label{fig:mappe-sigma}
\end{figure*}

Integration of these profiles yields the  stellar mass in each component, which turns
out to  be 52\%  in MR stars  and 48\%  in MP  stars. Although the  presence of  a MP
component in the bulge has been established beyond any doubt by all the spectroscopic
studies reviewed in the introduction, this is the first time that its contribution to
the bulge stellar mass budget has been evaluated.

\subsection{Comparison between mass and velocity dispersion}
Something interesting to explore is the  comparison between the stellar mass profile,
directly  inferred from  the  number of  stars, and  the  radial velocity  dispersion
profile.   The  latter is  widely  used  as tracer  of  the  total (dynamical)  mass,
especially  in   extragalactic  context,  modulo   some  assuption  on   the  spatial
distribution of the  stars (usually requiring fit of the  surface brightness profile)
and the distribution of the orbits.

A map  of the  global galactocentric radial  velocity ($\sigma$), for  the MP  and MR
components together, was published  in \citet[their Fig.~11]{zoccali+14}. Noticeably,
the map shows a central $\sigma$-peak, reaching $\sim$140 km/s, that was not expected
based    on    previous    measurements    in    more    external    bulge    regions
\citep[e.g.,][]{howard+08,freeman+13}.  The presence  of the  peak was  rather firmly
established by  the GIBS  data, as it  resulted from measurement  of the  $\sigma$ in
three independent fields,  each sampling 441, 435 and 111  stars. However its spatial
extension and symmetry  about the Galactic plane was poorly  constrained, since the 3
GIBS fields were rather close to each other, and all at negative latitudes. In a more
recent paper, Valenti et  al. 2018 (A\&A, {\it in press}), used  MUSE data to measure
the $\sigma$ in four  fields, two of which at $b$=$+2^\circ$,  and confirmed both the
presence and the symmetry of the peak.

Values for $\sigma$ of the MP and MR components, individually, have been measured
by \citet{zoccali+17}, in each of the GIBS  fields (c.f., their Fig.~11 and 12). 
 By using the {\it akima} package in R (Akima 1978), we applied a bivariate linear 
interpolation to the irregular grid of 26 GIBS fields after imposing a 4-fold symmetry
and derive spatial maps of sigma for MP and MR stars shown in Fig.\ref{fig:mappe-sigma}.
We  recall that this interpolation
allow us to  examine the run  of a variable ($\sigma$  in this
case) in  strips at constant  latitude, which would be  impossible with the  raw data
because -- in order to minimize extinction -- the observed fields were not exactly on
a regular grid. Second, the interpolation allows  us to predict the value of $\sigma$
at any position intermediate between our  fields. Third, it allows a straightforward
comparison with IFU measurements of external, edge-on galaxies \citep{gonzalez+16}.

The right panels  of Fig.~\ref{fig:mass-sigma} show the trend of  $\sigma$ for the MP
(blue)  and the  MR  (red) stars,  as  a  function of  longitude,  at three  constant
latitudes (the same  as for the mass  trends on the left  panels). Interestingly, the
$\sigma$ has an opposite trend with respect to the mass (or to the stellar density).
Namely, at intermediate  heights from the Galactic plane, MR  stars dominate the star
counts,  especially close  to the  projected  minor axis  ($l$=$0$). Their  $\sigma$,
however, is significantly lower than that of  MP stars. This behaviour is reverted at
$b$=$\pm1$, where MP  stars dominate the star counts, hence  the stellar mass budget,
but  their  $\sigma$  is   lower  than  that  of  MR  stars.
 In other words, at ($l,b$)=($0,-1$), MR stars have a significantly larger
velocity dispersion, that does not correspond to a significantly larger
contribution to the mass. This is because MR stars, arranged in a bar, have mostly
elongated orbits, which increase the observed 1-D sigma. Curiously, 
this behaviour ir inverted at b=+/-5, where MP stars have a larger sigma
although their contribution to the mass budget is lower.

\begin{table}
  \caption{Percent contribution of MP and MR stars to the bulge stellar mass as a
    function of Galactic position.
\label{tab:mass} }
\centering
\begin{tabular}{cccc}
\hline\hline
$l$ & $b$ & \% MP & \% MR\\
\hline
$-9$ & $-9$ & 0.0138 & 0.0100 \\
$-9$ & $-8$ & 0.0171 & 0.0143 \\
$-9$ & $-7$ & 0.0228 & 0.0200 \\
$-9$ & $-6$ & 0.0318 & 0.0280 \\
 .... &.... &.... &.... \\
\hline\hline
&&&\\
\multicolumn{4}{l}{{\bf Note}: The full table is available online at CDS.}\\
\end{tabular}
\end{table}

\section{Conclusions}

With  the increasing  amount  of photometric  and spectroscopic  data  over the  last
decade, the  complexity of the  Galactic bulge morphological, dynamical,  and stellar
population properties has  become evident. In particular, one of  the properties that
has been largely discussed  in the literature is  the different behaviour of  MP and MR
bulge stars.

Although from  a purely  observational perspective,  it is  not possible  to separate
bulge populations on the basis of a  given formation mechanism, in this study we used
the fact that the metallicity distribution function  can be separated in at least two
populations in  a statistically significant  way. The separation  between populations
are consistently seen across the bulge  and their variation can be quantified without
any assumptions on  their physical origin. Adequate numerical  models and simulations
are now available via which the characteristics of these two population can be linked
to a partciular  bulge formation process. However, one missing  ingredient so far has
been the relative contribution of the MP  and MR populations to the total mass budget
of the bulge.

In this study we derive this number for the first time by using two results presented
recently by the  GIBS and VVV surveys, namely the  relative density distribution maps
of  MP  and MR  from  \citep{zoccali+17}  and the  mass  profile  of the  bulge  from
\citet{valenti+16}.   We  converted the  GIBS  stellar  density  maps  of MP  and  MR
populations into maps of stellar mass and integrate them across the entire bulge area
to derive their relative contribution to the total stellar mass of the bulge. We find
that the relative contribution to the total bulge  mass comes in a 52\% from MR stars
and 48\% from MP stars.

We also discuss the spatial variations of  both the mass and radial velocity dispersion
profiles of  MP and  MR components.  We show that  while MR  stars dominate  the star
counts at intermediate latitudes ($b$$\sim$$-5$) their $\sigma$ is considerably lower
than that of MP stars. This behaviour is reverted closer to the plane ($b$$\sim$$-1$)
where MP stars are dominant but their $\sigma$ is lower than that of MR stars.

We note that the connection between the velocity dispersion and the mass
  is not direct, as it involves the orbit anisotropy and the spatial distribution of the stars.
  It is not the purpose  of the present
paper to infer the physical connection between  the two.  We here focus on presenting
the data, emphasizing that they show complex trends that require detailed modeling to
be understood. Proper motions are available from VVV in the $|b|$$<$3 region, and -very
soon- from  Gaia for  the rest of  the bulge, for  the same  stars for which  we have
radial velocities  and metallicities,  therefore in the  near future  bulge formation
models  can  be  constrained  much  better than  before,  hopefully  allowing  us  to
definitely discard some scenarios in favor of the others.


\begin{acknowledgements}
MZ acknowledges support from the Ministry for the Economy, Development, and Tourism's
Programa  Iniciativa  Cient\'\i  fica  Milenio through  grant  IC120009,  awarded  to
Millenium Institute of Astrophysics (MAS), the BASAL CATA Center for Astrophysics and
Associated Technologies through grant PFB-06,  and from FONDECYT Regular 1150345. OAG
gratefully acknowledges the European Southern Observatory in Chile where part of this
work was completed under the Scientific Visitors Programme.
\end{acknowledgements}

\bibliographystyle{aa}
\bibliography{mybiblio}

\begin{thebibliography}{44}
\expandafter\ifx\csname natexlab\endcsname\relax\def\natexlab#1{#1}\fi

\bibitem[{{Athanassoula}(2005)}]{athanassoula05}
{Athanassoula}, E. 2005, \mnras, 358, 1477

\bibitem[{{Babusiaux}(2016)}]{babusiaux16}
{Babusiaux}, C. 2016, \pasa, 33, e026

\bibitem[{{Babusiaux} {et~al.}(2010){Babusiaux}, {G{\'o}mez}, {Hill}, {Royer},
  {Zoccali}, {Arenou}, {Fux}, {Lecureur}, {Schultheis}, {Barbuy}, {Minniti}, \&
  {Ortolani}}]{babusiaux+10}
{Babusiaux}, C., {G{\'o}mez}, A., {Hill}, V., {et~al.} 2010, \aap, 519, A77

\bibitem[{{Bensby} {et~al.}(2017){Bensby}, {Feltzing}, {Gould}, {Yee},
  {Johnson}, {Asplund}, {Mel{\'e}ndez}, {Lucatello}, {Howes}, {McWilliam},
  {Udalski}, {Szyma{\'n}ski}, {Soszy{\'n}ski}, {Poleski}, {Wyrzykowski},
  {Ulaczyk}, {Koz{\l}owski}, {Pietrukowicz}, {Skowron}, {Mr{\'o}z}, {Pawlak},
  {Abe}, {Asakura}, {Bhattacharya}, {Bond}, {Bennett}, {Hirao}, {Nagakane},
  {Koshimoto}, {Sumi}, {Suzuki}, \& {Tristram}}]{bensby+17}
{Bensby}, T., {Feltzing}, S., {Gould}, A., {et~al.} 2017, \aap, 605, A89

\bibitem[{{Calamida} {et~al.}(2015){Calamida}, {Sahu}, {Casertano}, {Anderson},
  {Cassisi}, {Gennaro}, {Cignoni}, {Brown}, {Kains}, {Ferguson}, {Livio},
  {Bond}, {Buonanno}, {Clarkson}, {Ferraro}, {Pietrinferni}, {Salaris}, \&
  {Valenti}}]{calamida+15}
{Calamida}, A., {Sahu}, K.~C., {Casertano}, S., {et~al.} 2015, \apj, 810, 8

\bibitem[{{Clarkson} {et~al.}(2008){Clarkson}, {Sahu}, {Anderson}, {Smith},
  {Brown}, {Rich}, {Casertano}, {Bond}, {Livio}, {Minniti}, {Panagia},
  {Renzini}, {Valenti}, \& {Zoccali}}]{clarkson+08}
{Clarkson}, W., {Sahu}, K., {Anderson}, J., {et~al.} 2008, \apj, 684, 1110

\bibitem[{{Debattista} {et~al.}(2017){Debattista}, {Ness}, {Gonzalez},
  {Freeman}, {Zoccali}, \& {Minniti}}]{debattista+17}
{Debattista}, V.~P., {Ness}, M., {Gonzalez}, O.~A., {et~al.} 2017, \mnras, 469,
  1587

\bibitem[{{D{\'e}k{\'a}ny} {et~al.}(2013){D{\'e}k{\'a}ny}, {Minniti},
  {Catelan}, {Zoccali}, {Saito}, {Hempel}, \& {Gonzalez}}]{dekany+13}
{D{\'e}k{\'a}ny}, I., {Minniti}, D., {Catelan}, M., {et~al.} 2013, \apjl, 776,
  L19

\bibitem[{{Freeman} {et~al.}(2013){Freeman}, {Ness}, {Wylie-de-Boer},
  {Athanassoula}, {Bland-Hawthorn}, {Asplund}, {Lewis}, {Yong}, {Lane}, {Kiss},
  \& {Ibata}}]{freeman+13}
{Freeman}, K., {Ness}, M., {Wylie-de-Boer}, E., {et~al.} 2013, \mnras, 428,
  3660

\bibitem[{{Gilmore} {et~al.}(2012){Gilmore}, {Randich}, {Asplund}, {Binney},
  {Bonifacio}, {Drew}, {Feltzing}, {Ferguson}, {Jeffries}, {Micela}, \&
  et~al.}]{gilmore+12}
{Gilmore}, G., {Randich}, S., {Asplund}, M., {et~al.} 2012, The Messenger, 147,
  25

\bibitem[{{Gonzalez} {et~al.}(2016){Gonzalez}, {Gadotti}, {Debattista},
  {Rejkuba}, {Valenti}, {Zoccali}, {Coccato}, {Minniti}, \&
  {Ness}}]{gonzalez+16}
{Gonzalez}, O.~A., {Gadotti}, D.~A., {Debattista}, V.~P., {et~al.} 2016, \aap,
  591, A7

\bibitem[{{Gonzalez} {et~al.}(2012){Gonzalez}, {Rejkuba}, {Zoccali}, {Valenti},
  {Minniti}, {Schultheis}, {Tobar}, \& {Chen}}]{gonzalez+12}
{Gonzalez}, O.~A., {Rejkuba}, M., {Zoccali}, M., {et~al.} 2012, \aap, 543, A13

\bibitem[{{Gran} {et~al.}(2016){Gran}, {Minniti}, {Saito}, {Zoccali},
  {Gonzalez}, {Navarrete}, {Catelan}, {Contreras Ramos}, {Elorrieta},
  {Eyheramendy}, \& {Jord{\'a}n}}]{gran+16}
{Gran}, F., {Minniti}, D., {Saito}, R.~K., {et~al.} 2016, \aap, 591, A145

\bibitem[{{Haywood} {et~al.}(2016){Haywood}, {Di Matteo}, {Snaith}, \&
  {Calamida}}]{haywood+16}
{Haywood}, M., {Di Matteo}, P., {Snaith}, O., \& {Calamida}, A. 2016, \aap,
  593, A82

\bibitem[{{Hill} {et~al.}(2011){Hill}, {Lecureur}, {Gomez}, {Zoccali},
  {Schultheis}, {Babusiaux}, {Royer}, {Barbuy}, {Arenou}, {Minniti}, \&
  {Ortolani}}]{hill+11}
{Hill}, V., {Lecureur}, A., {Gomez}, A., {et~al.} 2011, \aap
  [\eprint[arXiv]{1107.5199}]

\bibitem[{{Howard} {et~al.}(2008){Howard}, {Rich}, {Reitzel}, {Koch}, {De
  Propris}, \& {Zhao}}]{howard+08}
{Howard}, C.~D., {Rich}, R.~M., {Reitzel}, D.~B., {et~al.} 2008, \apj, 688,
  1060

\bibitem[{{Kunder} {et~al.}(2016){Kunder}, {Rich}, {Koch}, {Storm}, {Nataf},
  {De Propris}, {Walker}, {Bono}, {Johnson}, {Shen}, \& {Li}}]{kunder+16}
{Kunder}, A., {Rich}, R.~M., {Koch}, A., {et~al.} 2016, \apjl, 821, L25

\bibitem[{{McWilliam} \& {Zoccali}(2010)}]{mwz10}
{McWilliam}, A. \& {Zoccali}, M. 2010, \apj, 724, 1491

\bibitem[{{Minniti} {et~al.}(2010){Minniti}, {Lucas}, {Emerson}, {Saito},
  {Hempel}, {Pietrukowicz}, {Ahumada}, {Alonso}, {Alonso-Garcia}, {Arias},
  {Bandyopadhyay}, {Barb{\'a}}, {Barbuy}, {Bedin}, {Bica}, {Borissova},
  {Bronfman}, {Carraro}, {Catelan}, {Clari{\'a}}, {Cross}, {de Grijs},
  {D{\'e}k{\'a}ny}, {Drew}, {Fari{\~n}a}, {Feinstein}, {Fern{\'a}ndez
  Laj{\'u}s}, {Gamen}, {Geisler}, {Gieren}, {Goldman}, {Gonzalez}, {Gunthardt},
  {Gurovich}, {Hambly}, {Irwin}, {Ivanov}, {Jord{\'a}n}, {Kerins}, {Kinemuchi},
  {Kurtev}, {L{\'o}pez-Corredoira}, {Maccarone}, {Masetti}, {Merlo},
  {Messineo}, {Mirabel}, {Monaco}, {Morelli}, {Padilla}, {Palma}, {Parisi},
  {Pignata}, {Rejkuba}, {Roman-Lopes}, {Sale}, {Schreiber}, {Schr{\"o}der},
  {Smith}, {}, {Soto}, {Tamura}, {Tappert}, {Thompson}, {Toledo}, {Zoccali}, \&
  {Pietrzynski}}]{minniti+10}
{Minniti}, D., {Lucas}, P.~W., {Emerson}, J.~P., {et~al.} 2010, \na, 15, 433

\bibitem[{{Nataf}(2016)}]{nataf+16}
{Nataf}, D.~M. 2016, \pasa, 33, e023

\bibitem[{{Nataf} {et~al.}(2010){Nataf}, {Udalski}, {Gould}, {Fouqu{\'e}}, \&
  {Stanek}}]{nataf+10}
{Nataf}, D.~M., {Udalski}, A., {Gould}, A., {Fouqu{\'e}}, P., \& {Stanek},
  K.~Z. 2010, \apjl, 721, L28

\bibitem[{{Ness} {et~al.}(2013{\natexlab{a}}){Ness}, {Freeman}, {Athanassoula},
  {Wylie-de-Boer}, {Bland-Hawthorn}, {Asplund}, {Lewis}, {Yong}, {Lane}, \&
  {Kiss}}]{ness+13a}
{Ness}, M., {Freeman}, K., {Athanassoula}, E., {et~al.} 2013{\natexlab{a}},
  \mnras, 430, 836

\bibitem[{{Ness} {et~al.}(2013{\natexlab{b}}){Ness}, {Freeman}, {Athanassoula},
  {Wylie-de-Boer}, {Bland-Hawthorn}, {Asplund}, {Lewis}, {Yong}, {Lane},
  {Kiss}, \& {Ibata}}]{ness+13b}
{Ness}, M., {Freeman}, K., {Athanassoula}, E., {et~al.} 2013{\natexlab{b}},
  \mnras, 432, 2092

\bibitem[{{Ness} {et~al.}(2012){Ness}, {Freeman}, {Athanassoula},
  {Wylie-De-Boer}, {Bland-Hawthorn}, {Lewis}, {Yong}, {Asplund}, {Lane},
  {Kiss}, \& {Ibata}}]{ness+12}
{Ness}, M., {Freeman}, K., {Athanassoula}, E., {et~al.} 2012, \apj, 756, 22

\bibitem[{{Ortolani} {et~al.}(1995){Ortolani}, {Renzini}, {Gilmozzi},
  {Marconi}, {Barbuy}, {Bica}, \& {Rich}}]{ortolani+95}
{Ortolani}, S., {Renzini}, A., {Gilmozzi}, R., {et~al.} 1995, \nat, 377, 701

\bibitem[{{Patsis} {et~al.}(2002){Patsis}, {Skokos}, \&
  {Athanassoula}}]{patsis+02}
{Patsis}, P.~A., {Skokos}, C., \& {Athanassoula}, E. 2002, \mnras, 337, 578

\bibitem[{{Pietrukowicz} {et~al.}(2015){Pietrukowicz}, {Koz{\l}owski},
  {Skowron}, {Soszy{\'n}ski}, {Udalski}, {Poleski}, {Wyrzykowski},
  {Szyma{\'n}ski}, {Pietrzy{\'n}ski}, {Ulaczyk}, {Mr{\'o}z}, {Skowron}, \&
  {Kubiak}}]{pietrukowicz+15}
{Pietrukowicz}, P., {Koz{\l}owski}, S., {Skowron}, J., {et~al.} 2015, \apj,
  811, 113

\bibitem[{{Pietrukowicz} {et~al.}(2012){Pietrukowicz}, {Udalski},
  {Soszy{\'n}ski}, {Nataf}, {Wyrzykowski}, {Poleski}, {Koz{\l}owski},
  {Szyma{\'n}ski}, {Kubiak}, {Pietrzy{\'n}ski}, \& {Ulaczyk}}]{pietrukowicz+12}
{Pietrukowicz}, P., {Udalski}, A., {Soszy{\'n}ski}, I., {et~al.} 2012, \apj,
  750, 169

\bibitem[{{Rojas-Arriagada} {et~al.}(2017){Rojas-Arriagada}, {Recio-Blanco},
  {de Laverny}, {Mikolaitis}, {Matteucci}, {Spitoni}, {Schultheis}, {Hayden},
  {Hill}, {Zoccali}, {Minniti}, {Gonzalez}, {Gilmore}, {Randich}, {Feltzing},
  {Alfaro}, {Babusiaux}, {Bensby}, {Bragaglia}, {Flaccomio}, {Koposov},
  {Pancino}, {Bayo}, {Carraro}, {Casey}, {Costado}, {Damiani}, {Donati},
  {Franciosini}, {Hourihane}, {Jofr{\'e}}, {Lardo}, {Lewis}, {Lind}, {Magrini},
  {Morbidelli}, {Sacco}, {Worley}, \& {Zaggia}}]{rojas-arriagada+17}
{Rojas-Arriagada}, A., {Recio-Blanco}, A., {de Laverny}, P., {et~al.} 2017,
  \aap, 601, A140

\bibitem[{{Rojas-Arriagada} {et~al.}(2014){Rojas-Arriagada}, {Recio-Blanco},
  {Hill}, {de Laverny}, {Schultheis}, {Babusiaux}, {Zoccali}, {Minniti},
  {Gonzalez}, {Feltzing}, {Gilmore}, {Randich}, {Vallenari}, {Alfaro},
  {Bensby}, {Bragaglia}, {Flaccomio}, {Lanzafame}, {Pancino}, {Smiljanic},
  {Bergemann}, {Costado}, {Damiani}, {Hourihane}, {Jofr{\'e}}, {Lardo},
  {Magrini}, {Maiorca}, {Morbidelli}, {Sbordone}, {Worley}, {Zaggia}, \&
  {Wyse}}]{rojas-arriagada+14}
{Rojas-Arriagada}, A., {Recio-Blanco}, A., {Hill}, V., {et~al.} 2014, \aap,
  569, A103

\bibitem[{{Saito} {et~al.}(2012){Saito}, {Hempel}, {Minniti}, {Lucas},
  {Rejkuba}, {Toledo}, {Gonzalez}, {Alonso-Garc{\'{\i}}a}, {Irwin},
  {Gonzalez-Solares}, {Hodgkin}, {Lewis}, {Cross}, {Ivanov}, {Kerins},
  {Emerson}, {Soto}, {Am{\^o}res}, {Gurovich}, {D{\'e}k{\'a}ny}, {Angeloni},
  {Beamin}, {Catelan}, {Padilla}, {Zoccali}, {Pietrukowicz}, {Moni Bidin},
  {Mauro}, {Geisler}, {Folkes}, {Sale}, {Borissova}, {Kurtev}, {Ahumada},
  {Alonso}, {Adamson}, {Arias}, {Bandyopadhyay}, {Barb{\'a}}, {Barbuy},
  {Baume}, {Bedin}, {Bellini}, {Benjamin}, {Bica}, {Bonatto}, {Bronfman},
  {Carraro}, {Chen{\`e}}, {Clari{\'a}}, {Clarke}, {Contreras}, {Corvill{\'o}n},
  {de Grijs}, {Dias}, {Drew}, {Fari{\~n}a}, {Feinstein},
  {Fern{\'a}ndez-Laj{\'u}s}, {Gamen}, {Gieren}, {Goldman},
  {Gonz{\'a}lez-Fern{\'a}ndez}, {Grand}, {Gunthardt}, {Hambly}, {Hanson},
  {He{\l}miniak}, {Hoare}, {Huckvale}, {Jord{\'a}n}, {Kinemuchi}, {Longmore},
  {L{\'o}pez-Corredoira}, {Maccarone}, {Majaess}, {Mart{\'{\i}}n}, {Masetti},
  {Mennickent}, {Mirabel}, {Monaco}, {Morelli}, {Motta}, {Palma}, {Parisi},
  {Parker}, {Pe{\~n}aloza}, {Pietrzy{\'n}ski}, {Pignata}, {Popescu}, {Read},
  {Rojas}, {Roman-Lopes}, {Ruiz}, {Saviane}, {Schreiber}, {Schr{\"o}der},
  {Sharma}, {Smith}, {Sodr{\'e}}, {Stead}, {Stephens}, {Tamura}, {Tappert},
  {Thompson}, {Valenti}, {Vanzi}, {Walton}, {Weidmann}, \&
  {Zijlstra}}]{saito+12}
{Saito}, R.~K., {Hempel}, M., {Minniti}, D., {et~al.} 2012, \aap, 537, A107

\bibitem[{{Saito} {et~al.}(2011){Saito}, {Zoccali}, {McWilliam}, {Minniti},
  {Gonzalez}, \& {Hill}}]{saito+11}
{Saito}, R.~K., {Zoccali}, M., {McWilliam}, A., {et~al.} 2011, \aj, 142, 76

\bibitem[{{Schultheis} {et~al.}(2017){Schultheis}, {Rojas-Arriagada},
  {Garc{\'{\i}}a P{\'e}rez}, {J{\"o}nsson}, {Hayden}, {Nandakumar}, {Cunha},
  {Allende Prieto}, {Holtzman}, {Beers}, {Bizyaev}, {Brinkmann}, {Carrera},
  {Cohen}, {Geisler}, {Hearty}, {Fernandez-Tricado}, {Maraston}, {Minnitti},
  {Nitschelm}, {Roman-Lopes}, {Schneider}, {Tang}, {Villanova}, {Zasowski}, \&
  {Majewski}}]{schultheis+17}
{Schultheis}, M., {Rojas-Arriagada}, A., {Garc{\'{\i}}a P{\'e}rez}, A.~E.,
  {et~al.} 2017, \aap, 600, A14

\bibitem[{{Soszy{\'n}ski} {et~al.}(2011){Soszy{\'n}ski}, {Dziembowski},
  {Udalski}, {Poleski}, {Szyma{\'n}ski}, {Kubiak}, {Pietrzy{\'n}ski},
  {Wyrzykowski}, {Ulaczyk}, {Koz{\l}owski}, \& {Pietrukowicz}}]{soszynski+11}
{Soszy{\'n}ski}, I., {Dziembowski}, W.~A., {Udalski}, A., {et~al.} 2011,
  \actaa, 61, 1

\bibitem[{{Sumi}(2004)}]{sumi04}
{Sumi}, T. 2004, \mnras, 349, 193

\bibitem[{{Valenti} {et~al.}(2016){Valenti}, {Zoccali}, {Gonzalez}, {Minniti},
  {Alonso-Garc{\'{\i}}a}, {Marchetti}, {Hempel}, {Renzini}, \&
  {Rejkuba}}]{valenti+16}
{Valenti}, E., {Zoccali}, M., {Gonzalez}, O.~A., {et~al.} 2016, \aap, 587, L6

\bibitem[{{Valenti} {et~al.}(2013){Valenti}, {Zoccali}, {Renzini}, {Brown},
  {Gonzalez}, {Minniti}, {Debattista}, \& {Mayer}}]{valenti+13}
{Valenti}, E., {Zoccali}, M., {Renzini}, A., {et~al.} 2013, \aap, 559, A98

\bibitem[{{Wegg} \& {Gerhard}(2013)}]{wegg+13}
{Wegg}, C. \& {Gerhard}, O. 2013, \mnras, 435, 1874

\bibitem[{{Zoccali} {et~al.}(2000){Zoccali}, {Cassisi}, {Frogel}, {Gould},
  {Ortolani}, {Renzini}, {Rich}, \& {Stephens}}]{zoccali+00}
{Zoccali}, M., {Cassisi}, S., {Frogel}, J.~A., {et~al.} 2000, \apj, 530, 418

\bibitem[{{Zoccali} {et~al.}(2014){Zoccali}, {Gonzalez}, {Vasquez}, {Hill},
  {Rejkuba}, {Valenti}, {Renzini}, {Rojas-Arriagada}, {Martinez-Valpuesta},
  {Babusiaux}, {Brown}, {Minniti}, \& {McWilliam}}]{zoccali+14}
{Zoccali}, M., {Gonzalez}, O.~A., {Vasquez}, S., {et~al.} 2014, \aap, 562, A66

\bibitem[{{Zoccali} {et~al.}(2008){Zoccali}, {Hill}, {Lecureur}, {Barbuy},
  {Renzini}, {Minniti}, {G{\'o}mez}, \& {Ortolani}}]{zoccali+08}
{Zoccali}, M., {Hill}, V., {Lecureur}, A., {et~al.} 2008, \aap, 486, 177

\bibitem[{Zoccali {et~al.}(2003)Zoccali, Renzini, Ortolani, Greggio, Saviane,
  Cassisi, Rejkuba, Barbuy, Rich, \& Bica}]{zoccali+03}
Zoccali, M., Renzini, A., Ortolani, S., {et~al.} 2003, A\&A, 399, 931

\bibitem[{{Zoccali} \& {Valenti}(2016)}]{zoccali+16}
{Zoccali}, M. \& {Valenti}, E. 2016, \pasa, 33, e025

\bibitem[{{Zoccali} {et~al.}(2017){Zoccali}, {Vasquez}, {Gonzalez}, {Valenti},
  {Rojas-Arriagada}, {Minniti}, {Rejkuba}, {Minniti}, {McWilliam}, {Babusiaux},
  {Hill}, \& {Renzini}}]{zoccali+17}
{Zoccali}, M., {Vasquez}, S., {Gonzalez}, O.~A., {et~al.} 2017, \aap, 599, A12

\end{thebibliography}
\end{document}